\documentclass[conference, times]{IEEEtran}
\usepackage{float}
\usepackage{tabularx}
\usepackage{mathtools}
\usepackage{amsmath}
\usepackage{hhline}
\usepackage{silence}
\usepackage{graphicx}
\usepackage{soul}
\usepackage{subcaption}
\usepackage{xcolor}
\usepackage{multirow}
\usepackage{graphicx}
\usepackage{amsmath}
\usepackage[english]{babel}
\usepackage{booktabs}
\usepackage{array}
\usepackage{paralist}
\usepackage{threeparttable}
\usepackage{lipsum}
\usepackage{flushend}
\usepackage{cuted}
\usepackage{algorithm}
\usepackage{algorithmic}
\usepackage{amssymb}
\usepackage{multicol}

\usepackage{cite}
\usepackage{makecell}
\usepackage{url}
\usepackage[table]{xcolor}
\usepackage[normalem]{ulem}
\usepackage{colortbl}
\usepackage[hidelinks]{hyperref}

\usepackage{enumitem}

\newcommand{\ignore}[1]{}
\newcommand{\red}[1]{\textcolor{red}{#1}}

\newcommand{\green}[1]{{\textcolor{teal}{#1}}}

\usepackage{tikz}
\usepackage{balance}

\usepackage{listings}
\lstdefinestyle{promptstyle}{
  basicstyle=\ttfamily\footnotesize,   
  columns=fullflexible,
  keepspaces=true,
  breaklines=true,
  showstringspaces=false,
  frame=single,
  rulecolor=\color{promptgray},
  backgroundcolor=\color{yellow!10},
  xleftmargin=0.15em,
  xrightmargin=0.15em,
  aboveskip=0.15em,
  belowskip=0.15em,
  commentstyle=\color{promptgray},
  keywordstyle=\color{promptgray},
  stringstyle=\color{promptgray},
  escapeinside={(*@}{@*)} 
}

\definecolor{promptgray}{rgb}{0.5,0.5,0.5}
\definecolor{lightcream}{RGB}{255,250,235}

\definecolor{lightblue}{RGB}{242,247,252}

\definecolor{lightgray}{RGB}{248,248,248}

\definecolor{lightgreen}{RGB}{243,250,245}

\definecolor{lightpeach}{RGB}{255,246,240}

\definecolor{lightlavender}{RGB}{248,246,252}

\definecolor{lightcyan}{RGB}{242,250,250}

\definecolor{lightyellow}{RGB}{255,252,235}

\definecolor{gmidestimator_color}{RGB}{255,234,234}

\definecolor{equationgenerator_color}{RGB}{248,255,240}

\definecolor{equationrepair_color}{RGB}{240, 247, 255}

\definecolor{iter1bg}{RGB}{255,242,238}   
\definecolor{iter2bg}{RGB}{255,249,225}   
\definecolor{iter3bg}{RGB}{237,249,241}   

\definecolor{iter1frame}{RGB}{190,70,60}
\definecolor{iter2frame}{RGB}{190,145,25}
\definecolor{iter3frame}{RGB}{45,135,75}

\definecolor{failred}{RGB}{190,35,35}
\definecolor{passgreen}{RGB}{25,125,65}
\definecolor{darkblue}{RGB}{20,65,120}

\usepackage{adjustbox}

\usepackage{pifont}

\begin{document}
\pagestyle{empty}

\title{Route-Aware EM Surrogate for Automated \\LC Matching Network Optimization}
\author{
\IEEEauthorblockN{Subhadip Ghosh, Surya Srikar Peri, Ramesh Harjani, and Sachin S. Sapatnekar}
\IEEEauthorblockA{
\text{Department of Electrical and Computer Engineering, University of Minnesota, Minneapolis, MN, USA} 
}
}

\maketitle
\begin{abstract}
This paper proposes an open-source automated LC matching network optimization framework that combines schematic-level design optimization with electromagnetic (EM)-aware post-layout refinement. An ML-based surrogate predicts the frequency-dependent complex admittance of individual routes, enabling reuse across matching network topologies and physical layouts for the same technology node without retraining. These route models guide post-layout passive retuning to recover matching performance while preserving the layout footprint. Evaluated in a commercial 65\,nm CMOS technology, the surrogate achieves 1.94--4.91\% mean absolute percentage error with respect to a full-wave electromagnetic solver, with no detected passivity violations. Across nine LC matching networks operating at 5--95\,GHz, our framework recovers $S_{11}$ from $-$6.37\,dB, for schematic-level optimization after post-layout degradation, 
to $-$22.27\,dB after EM-aware post-layout refinement. All designs meet matching targets under full-layout EM verification in optimization runtimes below 70\,s. 
\end{abstract}


\maketitle
\section{Introduction}
\label{sec:introduction}

\noindent
Modern wireless communication systems increasingly operate at radio-frequency (RF) and millimeter-wave (mm-wave) frequencies to support high data rates and wide bandwidths. At these frequencies, impedance matching is essential to reduce reflections and maximize power transfer between circuit blocks with different source ($Z_s$) and load ($Z_l$) impedances~\cite{pozar2011microwave,razavi2012rf}. Inductance-capacitance (LC) matching networks are widely used to realize this at a target operating frequency. Their design commonly begins with schematic-level optimization, which selects the topology and passive values to meet matching targets under ideal, lossless interconnect assumptions.

Prior work has explored schematic-level topology search and passive-value optimization~\cite{chen2012gasa,yarman}. To improve physical design accuracy, space mapping and surrogate-assisted EM optimization~\cite{bandler1994space,liu2014gaspad,liu2025egaspad}, automated passive synthesis and layout generation~\cite{ghosh2026composer}, and Gaussian-process/Bayesian optimization of matching network layouts~\cite{chen2026lc} have been investigated. Learned EM surrogates have also been developed for passive structures and matching networks~\cite{passos2023pacosyt,er2021transformer,lee2022transformer,he2026motif,karahan2023passives,chae2024pulserf,tian2026pixelated}. While these methods improve the design of passive and matching network modules, they do not explicitly account for the impact of the external routing interconnects connecting these modules to the rest of the system. Perfect matching at the schematic stage may translate to significant losses in a placed and routed design.


\begin{figure}[t]
    \centering
    \includegraphics[width=\columnwidth]{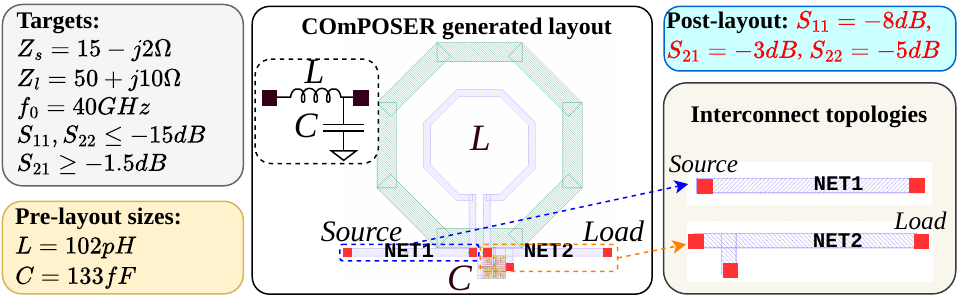}
    \caption{Performance degradation due to layout-oblivious matching network design and unmodeled routing EM effects.}
    \label{fig:overall_idea}
\end{figure}

Fig.~\ref{fig:overall_idea} illustrates this limitation using a schematic-level optimized matching network. After layout generation using COmPOSER~\cite{ghosh2026composer}, unmodeled routing parasitics degrade input/output reflection ($S_{11}$/$S_{22}$) and transmission ($S_{21}$), violating matching targets. Recovering these targets requires post-layout passive retuning, which can alter passive dimensions, terminal locations, and routing responses. Evaluating these changes requires costly EM simulations~\cite{xia2020cpw,aikio2021kaband}, potentially repeated across retuning iterations. 
Efficient retuning therefore requires a generalizable EM model that captures varying route geometries and terminal connectivities (e.g., point-to-point (P2P) routes (NET1) and branched multiport routes (NET2), as shown in Fig.~\ref{fig:overall_idea}) without repeated full EM simulations.

To address these requirements, this work proposes an automated matching network optimization framework using a route-level EM surrogate model. 
We train a multilayer perceptron (MLP) surrogate to predict the frequency-dependent complex admittance of P2P routes from their geometries. Passivity-aware training enforces passivity of the predicted admittances at the sampled frequency~\cite{gustavsen2001passivity,triverio2007consistency}. Unlike complete-network models, the route-level surrogate can be reused across matching topologies and layouts within its trained geometry, metal-stack, and technology domain.
Starting from the matching requirements, we use differential evolution (DE)~\cite{storn1997differential} to optimize passive values across candidate LC topologies to build the matching network at the schematic level, followed by physical layout generation using the open-source netlist-to-GDSII tool COmPOSER~\cite{ghosh2026composer}. We decompose branched multiport routes into P2P segments and reconstruct their admittances at the original terminals using the Kron reduction~\cite{dorfler2013kron}. 
These reconstructed route models guide post-layout passive retuning that updates the $LC$ values in the matching network based on routing parasitics,  preserving the selected topology and layout footprint to ensure convergence. We then generate the updated layout and verify its performance using full EM simulation. We evaluate the proposed framework across nine LC matching network designs with varying operating frequencies and bandwidths up to 100\,GHz in commercial 65\,nm CMOS technology. Post-layout retuning achieves up to 83.2\% power transfer and $S_{11}$ as low as $-27.5$\,dB within 70\,s, with all nine designs meeting matching targets under full EM verification.

\begin{figure}[t]
 \centering
 \includegraphics[width=\columnwidth]{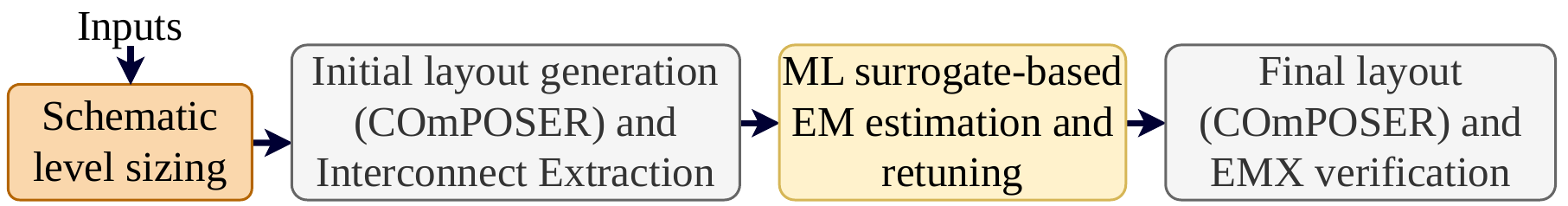}
 \caption{Proposed LC matching network design flow.}
 \label{fig:framework_flow}
\end{figure}

\section{Proposed Framework}
\label{sec:framework}

\subsection{Schematic-level sizing of matching networks}
\label{subsec:schematic_sizing}

\begin{figure}[t]
    \centering
    \includegraphics[width=0.9\columnwidth]{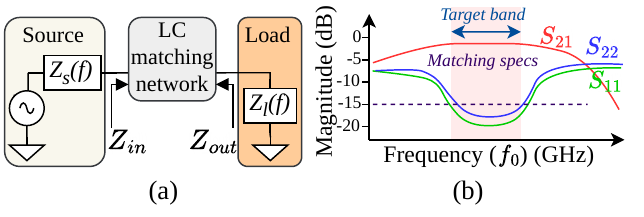}
    \caption{(a) LC matching problem and (b) desired S-parameter response over the target band.}
    \label{fig:matching_problem}
\end{figure}

\noindent
The proposed flow (Fig.~\ref{fig:framework_flow}) begins with schematic-level sizing of candidate LC matching networks. Inputs include the matching requirements (as illustrated by the targets in Fig.~\ref{fig:overall_idea}): $Z_s(f)$, $Z_l(f)$, $f_0$, LC topology library, and technology. Fig.~\ref{fig:matching_problem}(a) illustrates the source-to-load matching problem. Maximum power transfer ideally requires conjugate matching, $Z_{in}(f)=Z_s^*(f)$ and $Z_{out}(f)=Z_l^*(f)$, through optimized $LC$  values.
Fig.~\ref{fig:matching_problem}(b) shows the desired S-parameter
response over the target band, 
with $S_{11}$ and $S_{22}$ below the dotted specification line to minimize reflected power.
Meanwhile, $S_{21}$ should be
maximized to improve source-to-load power delivery. 

For each topology, we employ a DE~\cite{storn1997differential}-based approach to optimize $LC$ values using population-based candidate generation and cost-based selection. Each candidate is evaluated by stamping ideal passive admittances, $y_L=1/(j2\pi fL)$ and $y_C=j2\pi fC$, into a nodal matrix. Two nodal solves, one for an excitation at each port, provide the power-wave S-parameters referenced to $Z_s(f)$ and $Z_l(f)$. The matching cost, $F_{c}$, is evaluated over $N_f$ frequency samples as
\begin{equation}
F_c\!=\!\frac{1}{N_f}\sum_{k=1}^{N_f}\!
\big[|S_{11}(f_k)|^2+\max(0,1-|S_{21}(f_k)|^2)+|S_{22}(f_k)|^2\big]
\label{eq:sizing_objective}
\end{equation}
\noindent
These terms penalize input reflection, undelivered available power, and output reflection, respectively. For single-frequency matching, $N_f=1$ at $f_0$; otherwise, samples span the target band (Fig.~\ref{fig:matching_problem}(b)). Prescribed S-parameter limits must be obeyed at all sampled frequencies. The optimized topologies are ranked by cost, with fewer components preferred for ties. The selected design then proceeds to initial layout generation.

\subsection{Initial layout generation}
\label{subsec:initial_layout}

\noindent
As shown in Fig.~\ref{fig:framework_flow}, the selected topology and optimized $LC$ values are passed to COmPOSER~\cite{ghosh2026composer,composer}. COmPOSER maps passive values to PDK-realizable layouts, optimized for high quality factor and low loss, and performs placement and routing within the available layout region while maintaining fixed source/load interfaces. The resulting GDSII provides the interconnect geometries, passive footprints, and terminal locations for subsequent extraction and route-aware retuning.

\subsection{ML surrogate-based EM estimation and retuning}
\label{subsec:interconnect_ml}
\label{subsec:kron_retuning}

\noindent
Accurate estimation of interconnect parasitics is essential for layout-aware matching network optimization, but repeated EM simulations during retuning are computationally expensive. We therefore employ an MLP-based surrogate to predict the frequency-dependent admittance of routed interconnects from their geometry. Depending on their connectivity, interconnects may connect two or more terminals~(Fig.~\ref{fig:overall_idea}). P2P routes are modeled directly as 2-port networks, while multiport routes are decomposed into P2P segments whose predicted admittances are assembled and reconstructed using Kron reduction. This allows the same surrogate to model different terminal counts and routing configurations.

\noindent\textbf{Point-to-point routes.}
The surrogate is trained on 50,000 EMX-characterized, PDK-valid multilayer Manhattan P2P routes, generated in about 6\,hours. In a commercial 65\,nm CMOS, the upper metal layers (M8--M9) are used due to their low resistance and suitability for RF/mm-wave design. This choice is technology-specific; other technologies require appropriate low-resistivity layer selection and retraining. 
The training routes contain 0--3 bends and segment lengths of up to 100\,$\mu$m: such lengths are observed even in compact matching network layouts within LNA/PA designs, even though COmPOSER optimizes routes to favor short interconnects.
Each route uses 594 features describing terminal locations, metal polygons, segment dimensions, bends, and vias. Hyperparameter tuning selects a 3-hidden-layer MLP (128/256/128 neurons) to predict the complex 2-port admittance matrix
\begin{equation}
\widehat{Y}(f)=
\begin{bmatrix}
\widehat{Y}_{11} & \widehat{Y}_{12}\\
\widehat{Y}_{21} & \widehat{Y}_{22}
\end{bmatrix}
\end{equation}
at 200 frequency samples up to 100\,GHz, yielding 1600 outputs from the real/imaginary parts of four admittance entries.

Non-passive surrogate predictions can introduce spurious power generation and physically invalid retuning solutions. We ensure passivity at each sampled frequency through passivity-aware training, followed by correction of remaining violations before optimization. For the Hermitian part $H=(\widehat{Y}+\widehat{Y}^{H})/2$, passivity requires $H\succeq0$, i.e., all eigenvalues \{$\lambda_k$\} of $H$ must be nonnegative~\cite{gustavsen2001passivity,triverio2007consistency}. The training loss includes a penalty:
\begin{equation}
\begin{aligned}
\mathcal{L}_{\rm pass}
&=\underset{f,k}{\rm mean}
\left[\frac{\max(0,-\lambda_k(H)-\epsilon)}{s_f}\right]^2,\\
\mathcal{L}_{\rm train}
&=\mathcal{L}_{\rm MSE}+\alpha\mathcal{L}_{\rm pass},
\end{aligned}
\label{eq:passivity_loss}
\end{equation}
where $\epsilon=10^{-6}$\,S, $s_f$ scales the penalty, and $\alpha=10$ after a five-epoch warmup. This regularization reduces passivity violations in the ML model from 51.07\% to 0.26\% and MAE from 15.39\,mS to 4.23\,mS. Remaining violations are corrected by interpolating between adjacent passive frequency 
samples. Thus, all admittances are guaranteed to satisfy passivity.

\begin{figure}[t]
    \centering
    \includegraphics[width=0.9\columnwidth]{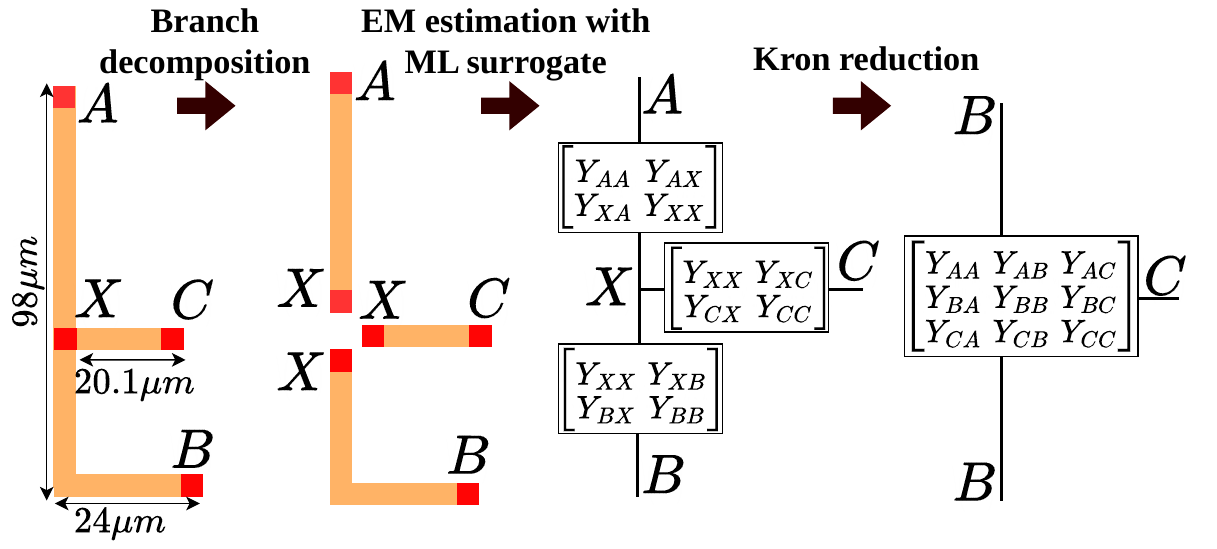}
    \caption{Point-to-point decomposition and Kron reconstruction of a branched multiport interconnect.}
    \label{fig:multiport_decomposition}
\end{figure}

\noindent\textbf{Multiport routes.}
For multiport routes such as NET2 in Fig.~\ref{fig:overall_idea}, the P2P surrogate is extended through decomposition and reconstruction. Fig.~\ref{fig:multiport_decomposition} illustrates decomposition of a branched route into P2P segments at internal junctions and reconstruction using Kron reduction. In the three-terminal example, the route connecting $A$, $B$, and $C$ through junction $X$ is decomposed into $A$--$X$, $X$--$B$, and $X$--$C$.
Each segment is evaluated using the same MLP surrogate, and its predicted 2-port admittance is stamped into a common nodal admittance matrix according to the connectivity~\cite{ho1975modified}.
With external nodes $e\!=\!\{\!A\!,\!B\!,\!C\!\}$ and internal node $i\!=\!\{\!X\!\}$, the assembled admittance matrix is partitioned into external ($Y_{ee}$), internal ($Y_{ii}$), and cross-coupling ($Y_{ei}$, $Y_{ie}$) blocks. Kron reduction~\cite{dorfler2013kron} eliminates the internal junction, yielding the reconstructed admittance:
$
Y_{\rm ext}(f)=
Y_{ee}(f)-Y_{ei}(f)Y_{ii}^{-1}(f)Y_{ie}(f),
\label{eq:kron}
$
at the original terminals. This avoids training separate surrogate models for different port count/branched interconnects.

\noindent\textbf{Interconnect-aware retuning.}
With the interconnect admittance models, the next stage incorporates these models into the matching network to compensate for the routing-induced performance degradation illustrated in Fig.~\ref{fig:overall_idea}. The objective is to recover the optimized schematic-level response by adjusting the $LC$ values while preserving the selected topology. The target admittance $Y_t(f)$ is obtained by assembling the optimized schematic-level passive admittances into a nodal matrix and eliminating internal nodes to obtain the two-port response at the source/load terminals. During retuning, $Y_t(f)$ remains fixed, while the routed network response $Y(f,\mathbf{x})$ is evaluated using current passive values $\mathbf{x}$ and reconstructed interconnect models.
The deviation from the target is expressed as
\begin{equation}
\mathbf{e}(\mathbf{x},f)=
\begin{bmatrix}
\Re\{Y(f,\mathbf{x})-Y_t(f)\}\\
\Im\{Y(f,\mathbf{x})-Y_t(f)\}
\end{bmatrix},
\label{eq:retune_error}
\end{equation}
stacking the real ($\Re$) and imaginary ($\Im$) errors at frequency $f$.

To avoid repeating global optimization, a first-order Taylor expansion approximates the error variation with passive values. The Jacobian $J$ captures the sensitivity of the error to each $LC$ value, giving $\mathbf{e}(\mathbf{x}+\Delta\mathbf{x})\approx\mathbf{e}(\mathbf{x})+J\Delta\mathbf{x}$. A bounded least-squares step determines the adjustments $\Delta\mathbf{x}$ that bring the routed response closer to $Y_t(f)$. 
The empirically chosen bounds $0.75\mathbf{x}_0\leq\mathbf{x}\leq1.25\mathbf{x}_0$ limit passive-value changes to $\pm25\%$ of the schematic solution $\mathbf{x}_0$ to help preserve the initial layout footprint and routing, avoiding repeated interconnect model updates due to changes in placement and routing.

Algorithm~\ref{alg:retuning} summarizes this iterative procedure. Retuning is initialized with the route models, target response, and schematic $LC$ values (Lines~\ref{line:retune_input}--\ref{line:retune_init}). 
Each iteration evaluates the routed response $Y(f,\mathbf{x})$~(Line~\ref{line:retune_response}) and its error (Line~\ref{line:retune_error}). 
Finite-difference perturbations of each $x_k$ estimate the Jacobian columns  $J_{:,k}$~(Lines~\ref{line:retune_perturb}--\ref{line:retune_jacobian}). Bounded least-squares updates of passive values continue until convergence (Lines~\ref{line:retune_solve}--\ref{line:retune_update}), yielding the retuned passive values (Line~\ref{line:retune_return}).

\begin{algorithm}[t]
\footnotesize
\caption{Route-Aware Passive Retuning}
\label{alg:retuning}
\setlength{\algorithmicindent}{0.6em}
\begin{algorithmic}[1]
\STATE \label{line:retune_input} \textbf{Input:} route models $Y_r(f)$, target $Y_t(f)$, initial $LC$ values $\mathbf{x}_0$
\STATE \label{line:retune_init} $\mathbf{x}\leftarrow\mathbf{x}_0$
\WHILE{not converged}
    \STATE \label{line:retune_response} Stamp current $LC$ values and $Y_r(f)$ to obtain $Y(f,\mathbf{x})$
    \STATE \label{line:retune_error} Compute $\mathbf{e}(\mathbf{x},f)$ using Eq.~\ref{eq:retune_error}
    \FOR{each passive value $x_k$}
        \STATE \label{line:retune_perturb} Perturb $x_k$ by $\Delta x_k$ and compute the resulting error $\mathbf{e}_{\rm pert}$
        \STATE \label{line:retune_jacobian} $J_{:,k}\leftarrow[\mathbf{e}_{\rm pert}-\mathbf{e}(\mathbf{x},f)]/\Delta x_k$
    \ENDFOR
    \STATE \label{line:retune_solve} $\Delta\mathbf{x}\leftarrow\arg\min\|\mathbf{e}(\mathbf{x},f)\!+\!J\Delta\mathbf{x}\|_2^2$ s.t. $0.75\mathbf{x}_0\leq\mathbf{x}+\Delta\mathbf{x}\leq1.25\mathbf{x}_0$
    \STATE \label{line:retune_update} $\mathbf{x}\leftarrow\mathbf{x}+\Delta\mathbf{x}$
\ENDWHILE
\STATE \label{line:retune_return} \textbf{return} retuned $LC$ values $\mathbf{x}$
\end{algorithmic}
\end{algorithm}

The returned values are mapped to realizable devices while preserving the existing layout footprint and source/load interfaces. Unchanged routes reuse surrogate predictions, while modified routes are regenerated and reevaluated. COmPOSER generates the updated layout, followed by full EMX verification of the complete matching network (Fig.~\ref{fig:framework_flow}).
\section{Experimental Setup and Results}
\label{sec:results}

\subsection{EM surrogate accuracy}
\label{subsec:em_accuracy}

\begin{table}[t]
\centering
\caption{2-port EM surrogate accuracy over 5,000 test routes.}
\label{tab:twoport_accuracy}
\scriptsize
\setlength{\tabcolsep}{4pt}
\renewcommand{\arraystretch}{0.9}
\arrayrulecolor{black}
\resizebox{\columnwidth}{!}{%
\begin{tabular}{l|cc||cc||cc||cc}
\hline
\rowcolor{gray!18}
&
\multicolumn{2}{c||}{\textbf{$Y_{11}$}} &
\multicolumn{2}{c||}{\textbf{$Y_{12}$}} &
\multicolumn{2}{c||}{\textbf{$Y_{21}$}} &
\multicolumn{2}{c}{\textbf{$Y_{22}$}} \\
\hhline{~--------}
\rowcolor{gray!18}
\multirow{-2}{*}{\textbf{Metric}} &
\textbf{Real} & \textbf{Imag.} &
\textbf{Real} & \textbf{Imag.} &
\textbf{Real} & \textbf{Imag.} &
\textbf{Real} & \textbf{Imag.} \\
\hline
MAE (mS) & 3.69 & 4.74 & 3.76 & 4.72 & 3.71 & 4.70 & 3.70 & 4.79 \\
MAPE (\%) & 3.09 & 2.07 & 2.99 & 1.94 & 2.99 & 1.94 & 3.14 & 2.09 \\
\hline
\end{tabular}%
}
\end{table}

\noindent
\textbf{P2P models.} We first evaluate the 2-port EM surrogate using 5,000 unseen routes (10\% of the dataset). The surrogate is trained for 50 epochs using AdamW with an adaptive learning rate initialized at $10^{-4}$. Table~\ref{tab:twoport_accuracy} reports mean absolute and percentage error (MAE and MAPE) for the real and imaginary parts of all four admittance parameters, averaged over test routes and frequency samples. The average MAE is 4.23\,mS across all eight outputs, with comparable accuracy for the transfer ($Y_{12}$, $Y_{21}$) and self-admittance ($Y_{11}$, $Y_{22}$) terms.


\begin{table}[t]
\centering
\caption{Multiport reconstruction accuracy compared to EMX.}
\vspace{-2mm}
\label{tab:results_multiport}
\scriptsize
\setlength{\tabcolsep}{1pt}
\renewcommand{\arraystretch}{0.9}
\arrayrulecolor{black}
\resizebox{\columnwidth}{!}{%
\begin{tabular}{l|cc||cc||cc}
\hline
\rowcolor{gray!18}
&
\multicolumn{2}{c||}{\textbf{3-port}} &
\multicolumn{2}{c||}{\textbf{4-port}} &
\multicolumn{2}{c}{\textbf{5-port}} \\
\rowcolor{gray!18}
\textbf{Reconstruction} &
\textbf{MAE (mS)} & \textbf{MAPE (\%)} &
\textbf{MAE (mS)} & \textbf{MAPE (\%)} &
\textbf{MAE (mS)} & \textbf{MAPE (\%)} \\
\hline
EMX segments + Kron & 12.43 & 3.45 & 19.73 & 4.06 & 21.15 & 4.78 \\
Surrogate segments + Kron & 13.01 & 4.83 & 18.92 & 4.13 & 22.03 & 4.91 \\
\hline
\end{tabular}%
}
\vspace{-2mm}
\end{table}

\begin{figure}[t]
    \centering
    \includegraphics[width=0.95\columnwidth]{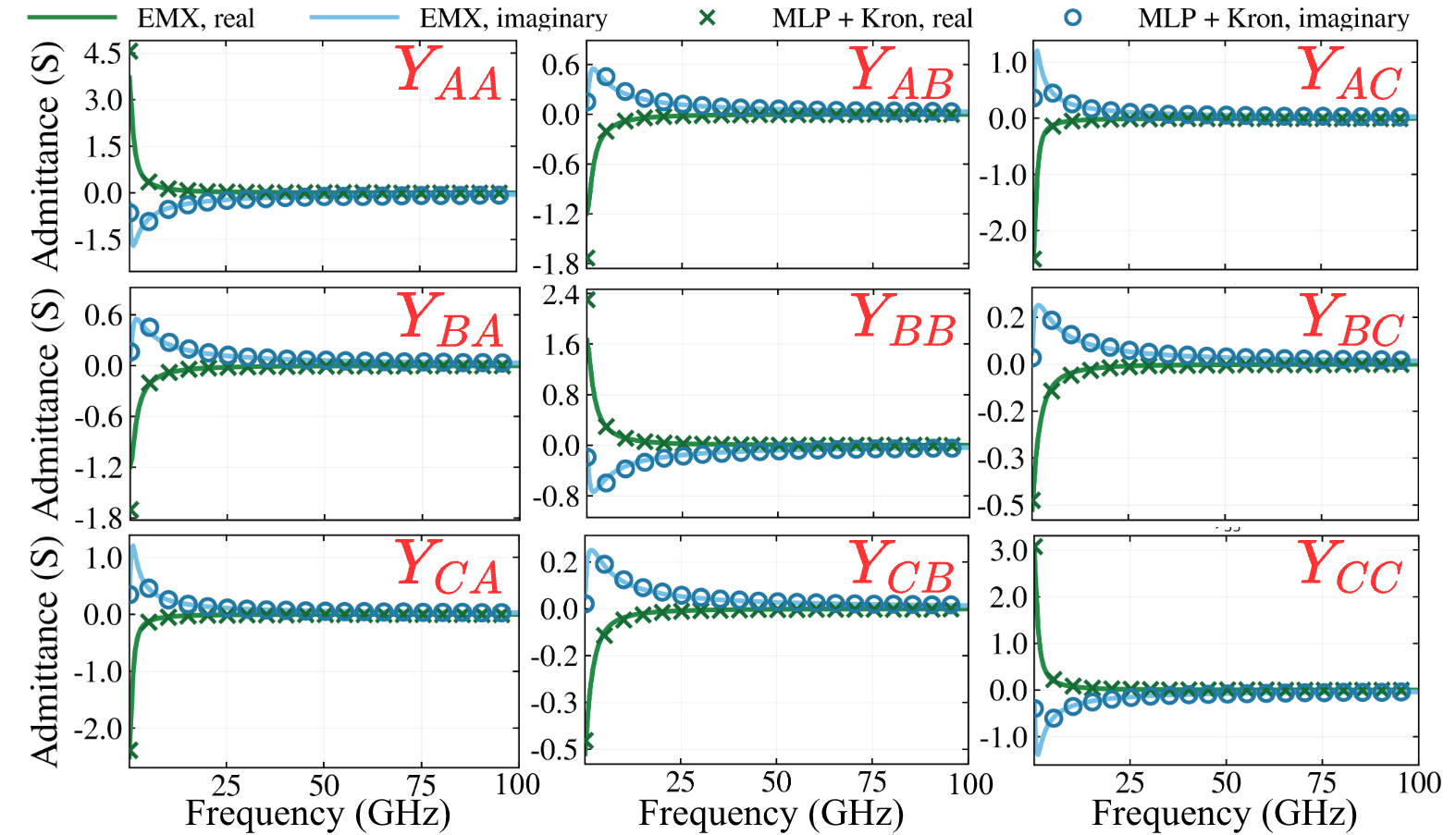}
    \vspace{-1mm}
    \caption{Full EMX vs. surrogate-based Kron-reconstructed admittance response for the 3-port route in Fig.~\ref{fig:multiport_decomposition}.}
    \label{fig:threeport_reconstruction}
    \vspace{-2mm}
\end{figure}

\noindent
\textbf{Multiport route model.} Next, we evaluate multiport reconstruction using the 2-port surrogate and Kron reduction (Section~\ref{subsec:kron_retuning}). We evaluate 50 unseen layouts each for three- to five-port nets, covering the maximum port count in our testcases. Full EMX simulation of each intact multiport net serves as the reference. Each net is reconstructed twice using the same Kron-reduction procedure: first from individually simulated EMX P2P segments and then from surrogate-predicted segments. The EMX-segment path isolates decomposition errors, primarily from neglected mutual EM coupling, while the surrogate path evaluates the complete reconstruction.

Table~\ref{tab:results_multiport} reports complex-admittance MAE and MAPE against EMX, averaged over layouts, admittance entries, and frequencies. Both reconstruction approaches exhibit comparable errors, with MAPE ranging from 3.45\% to 4.78\% for EMX segments and 4.13\% to 4.91\% for surrogate segments. The lower surrogate MAE for the 4-port case results from partial cancellation of prediction and decomposition errors. 
Comparing the two reconstructed admittances helps isolate the additional error from surrogate substitution.
Fig.~\ref{fig:threeport_reconstruction} further illustrates that the surrogate-based reconstruction closely tracks the EMX responses up to 100\,GHz across the $3\times3$ admittance matrix. These results support surrogate-based multiport reconstruction for route-aware retuning.

\begin{table}[t]
\centering
\caption{Evaluated testcases and their performance metrics \red{before} and \green{after} route-aware retuning. 
}
\label{tab:end_to_end}
\scriptsize
\setlength{\tabcolsep}{1pt}
\renewcommand{\arraystretch}{0.92}
\arrayrulecolor{black}
\resizebox{\columnwidth}{!}{%
\begin{tabular}{l|c|c|c|c|c|c|c|c|c|c}
\hline
\rowcolor{gray!18}
\textbf{Topology} &
\textbf{$N_p$} &
\textbf{$f_0$} &
\textbf{$Z_S$} &
\textbf{$Z_l$} &
\textbf{$f_c$} &
\textbf{$S_{11}$} &
\textbf{$S_{21}$} &
\textbf{$S_{22}$} &
\textbf{$G_T$} &
\textbf{Time} \\
\rowcolor{gray!18}
&
&
\textbf{(GHz)} &
\textbf{(Ohm)} &
\textbf{(Ohm)} &
\textbf{(GHz)} &
\textbf{(dB)} &
\textbf{(dB)} &
\textbf{(dB)} &
\textbf{(\%)} &
\textbf{(s)} \\
\hline
\multicolumn{11}{c}{\textbf{\textit{Single-frequency matching networks}}} \\
\hline
HP-L & 2 & 15 & 40+j10 & 60-j15 & \red{14.0}/\green{15.0} & \red{-7.1}/\green{-21.7} & \red{-3.0}/\green{-0.9} & \red{-2.1}/\green{-20.6} & \red{\textbf{50.1}}/\green{\textbf{81.2}} & 32.0 \\
LP-L & 2 & 40 & 45-j10 & 65+j15 & \red{35.7}/\green{40.0}& \red{-2.5}/\green{-22.2} & \red{-4.7}/\green{-0.8} & \red{-3.6}/\green{-20.2} & \red{\textbf{33.9}}/\green{\textbf{83.2}} & 33.5 \\
LP-$\pi$ & 3 & 65 & 35+j10 & 70-j15 & \red{54.9}/\green{65.0} & \red{-6.7}/\green{-19.7} & \red{-3.1}/\green{-1.1} & \red{-8.8}/\green{-18.9} & \red{\textbf{48.9}}/\green{\textbf{77.6}} & 36.9 \\
\hline
\multicolumn{11}{c}{\textbf{\textit{Narrowband matching networks}}} \\
\hline
HP-$\pi$ & 3 & 5--7 & 40+j10 & 60-j10 & \red{6.0}/\green{6.0} & \red{-6.0}/\green{-19.9} & \red{-5.1}/\green{-1.0} & \red{-7.1}/\green{-23.8} & \red{\textbf{30.9}}/\green{\textbf{79.4}} & 41.1 \\
BP-T & 6 & 38--42 & 45-j10 & 65+j10 & \red{36.5}/\green{40.0} & \red{-9.1}/\green{-24.5} & \red{-5.8}/\green{-1.2} & \red{-7.4}/\green{-24.1} & \red{\textbf{26.3}}/\green{\textbf{75.8}} & 51.8 \\
BP-$\pi$ & 6 & 89--95 & 35+j10 & 70-j15 & \red{79.0}/\green{91.0} & \red{-7.1}/\green{-27.5} & \red{-4.8}/\green{-0.9} & \red{-9.1}/\green{-24.6} & \red{\textbf{33.1}}/\green{\textbf{81.3}} & 61.4 \\
\hline
\multicolumn{11}{c}{\textbf{\textit{Wideband matching networks}}} \\
\hline
LP-5 & 5 & 35--42 & 40+j15 & 70-j10 & \red{34.0}/\green{38.5} & \red{-4.0}/\green{-21.4} & \red{-7.8}/\green{-1.1} & \red{-2.8}/\green{-20.9} & \red{\textbf{16.5}}/\green{\textbf{77.6}} & 55.7 \\
HP-5 & 5 & 67--80 & 35-j10 & 60+j15 & \red{68.0}/\green{73.5}& \red{-9.1}/\green{-22.0} & \red{-5.5}/\green{-0.9} & \red{-7.6}/\green{-20.7} & \red{\textbf{28.1}}/\green{\textbf{81.2}} & 64.1 \\
LP-7 & 7 & 80--92 & 45+j10 & 65-j10 & \red{74.0}/\green{86.0} & \red{-5.7}/\green{-21.5} & \red{-4.6}/\green{-1.3} & \red{-4.2}/\green{-19.8} & \red{\textbf{34.6}}/\green{\textbf{74.1}} & 68.9 \\
\hline
\end{tabular}%
}
\vspace{1pt}
\begin{minipage}{\columnwidth}
\footnotesize
\textbf{Notation:}
HP/LP/BP: high-/low-/band-pass;
L, T, $\pi$: network configurations;
$N_p$: passive-element count.
$A/B$: surrogate-estimated initial-layout /
full-EMX-verified post-retuning final-layout performance.
\end{minipage}
\end{table}

\subsection{End-to-end EM-aware matching network synthesis}
\label{subsec:mn_results}

\noindent
We evaluate the complete flow on nine matching problems: three single-frequency and six banded cases with bandwidths of 2--13\,GHz (Table~\ref{tab:end_to_end}). The framework and testcases are open-sourced at~\cite{testcases} and cover operating frequencies of 5--95\,GHz, varying complex source/load impedances ($Z_s$, $Z_l$), and networks with 2--7 passive elements. Single-frequency cases use compact L- and $\pi$-type networks at target frequencies, while banded cases employ $\pi$-, T-, and higher-order LC networks with additional tunable passive values to maintain matching across frequency bands.

For each test case, the matching-performance targets are $S_{11},S_{22}\leq -15$\,dB and $S_{21}\geq -1.5$\,dB. The reflection limits correspond to at most 3.2\% reflected power at either port. The percent forward power transfer $G_T=100\times10^{S_{21}(\mathrm{dB})/10}$~\cite{kurokawa1965power,pozar2011microwave}, with the $S_{21}$ target requiring at least 70\% transfer. The achieved center frequency, $f_c$, is also tracked against the single target frequency, $f_0$, or the band midpoint.
Single-frequency designs are evaluated at $f_0$; for banded designs, worst-case $S_{11}$ and $S_{22}$, and minimum $S_{21}$ and $G_T$ are reported over the target band, with final design compliance confirmed by EMX.

Table~\ref{tab:end_to_end} focuses on the performance recovered through route-aware retuning relative to the degraded initial physical layout, since schematic-level DE optimization ideally targets zero reflection and 100\% power transfer (Section~\ref{subsec:schematic_sizing}). Each $\red{A}/\green{B}$ entry for $f_c$, $S_{11}$, $S_{21}$, $S_{22}$, and $G_T$ denotes the \emph{surrogate-estimated initial-layout / full-EMX-verified post-retuning final-layout} performance. Thus, $\red{A}$ captures the degradation after physical realization, while $\green{B}$ shows the recovery achieved through route-aware retuning and layout regeneration. In the table, $N_p$ and Time denote the passive-element count and optimization runtime, respectively.

The initial layouts based on schematic optimization exhibit substantial matching degradation due to routing parasitics. Across nine cases, $G_T$ falls to 16.5--50.1\% ($S_{21}$: $-7.8$ to $-3.0$\,dB), while $S_{11}$ and $S_{22}$ degrade to $-9.1$ to $-2.5$\,dB and $-9.1$ to $-2.1$\,dB, respectively. No initial design satisfies the prescribed reflection or transmission targets; routing parasitics also shift $f_c$ from the target. The degradation is consistent with conductor losses and distributed EM effects of interconnect routes, ignored in the ideal schematic representation.

Route-aware retuning recovers the required performance in all nine cases, pulling the shifted center frequency $f_c$ back to its target value. Full-EMX verification of the regenerated layouts yields $S_{11}$ from $-27.5$ to $-19.7$\,dB, $S_{22}$ from $-24.6$ to $-18.9$\,dB, and $S_{21}$ from $-1.3$ to $-0.8$\,dB, corresponding to $G_T$ of 74.1--83.2\%. Even the worst final designs retain margins of 4.7\,dB for $S_{11}$, 3.9\,dB for $S_{22}$, and 0.2\,dB for $S_{21}$, while even the minimum $G_T$ exceeds the 70.8\% requirement by 3.3 percentage points. For banded cases, these values represent worst-case performance across the entire frequency interval.

The optimization runtime of 32.0--68.9\,s includes schematic-level DE optimization, COmPOSER layout generation, and route-aware post-layout retuning. 
On average, schematic-level DE optimization accounts for 3.5\% of the execution time, COmPOSER layout generation consumes 64.4\%, and route-aware post-layout retuning takes the remaining 32.1\%. Offline surrogate training and final EMX verification are excluded from this runtime 
(Section~\ref{subsec:interconnect_ml}). Overall, surrogate-guided retuning recovers the required matching performance across the evaluated topologies and operating frequencies, with final compliance independently verified using full-layout EMX.
\section{Conclusion}
\label{sec:conclusion}

\noindent
We present a route-aware LC matching network optimization framework using EM surrogates to model frequency-dependent interconnect parasitics. Post-layout retuning recovers matching performance lost during schematic-to-layout translation. The surrogate achieves 2.53\% average MAPE. Across nine matching network topologies spanning frequencies and bandwidths, retuning improves average power transfer by 45.4 percentage points and $S_{11}$ by 15.9\,dB, achieving 83.2\% power transfer and $S_{11}$ as low as $-27.5$\,dB within 70\,s. All meet targets under full-layout EM verification.

\clearpage
\balance
\bibliographystyle{misc/IEEEtran}
\bibliography{bib/main.bib}

\end{document}